\documentclass[manuscript]{aastex61}

\received{2018 August 24}
\revised{2019 July 29}
\accepted{2019 July 29}

\shorttitle{kappa-maxwellian electrons and bi-maxwellian protons}
\shortauthors{Taran, Safari, Daei}

\usepackage{amsmath}
\usepackage{graphicx}
\usepackage{xcolor}
\usepackage{lineno}
\begin{document}

\title{KAPPA-MAXWELLIAN ELECTRONS and BI-MAXWELLIAN PROTONS in a two-fluid model for fast solar wind}

\correspondingauthor{Hossein Safari}
\email{safari@znu.ac.ir}

\author{Somayeh Taran}
\affil{Department of Physics, Faculty of Science, University of Zanjan, P. O. Box 45195-313, Zanjan, Iran}

\author{Hossein Safari}
\affil{Department of Physics, Faculty of Science, University of Zanjan, P. O. Box 45195-313, Zanjan, Iran}

\author{Farhad Daei}
\affil{Department of Physics, Faculty of Science, University of Zanjan, P. O. Box 45195-313, Zanjan, Iran}
\begin{abstract}
Modeling fast solar wind based on the kinetic theory is an important task for scientists. In this paper, we present a two-fluid model for fast solar wind with anisotropic Kappa-Maxwellian electrons and Bi-Maxwellian  protons. In the simulation, the energy exchange between the plasma particles and low-frequency Alfv\'en waves is considered.
A set of eleven coupled equations is derived by applying the
zeroth- to fourth-order moments of the Vlasov equation and the modified 
electromagnetic Maxwell equations.
A characteristic of the Kappa distribution (indicated by $\kappa$ index) is explicit in the
equation for the parallel component of the electron heat flux (parallel to the ambient magnetic field line) and differs from the equation derived for the proton heat flux due to the different nature of the
distributions.
Within the large $\kappa$ index, the equations for the two-fluid model tend to the equations obtained by the Maxwellian distribution. Using an iterated Crank-Nicolson method, the coupled equations are numerically solved for the fast solar wind conditions.
 We show that at (0.3 - 1) AU from the Sun, the electron density, components of temperature, and  components of heat flux follow the power-law behavior. We also showed that near the Earth, the flow speed (electron or proton) increases with decreasing $\kappa$. We concluded that applying the small $\kappa$ index (the non-Maxwellian distribution), the extraordinary nature of the solar atmosphere, with its temperature of several million kelvin temperature for electrons, has been captured.
\end{abstract}

\section{Introduction}\label{sect1}
Cosmic rays and the huge volume of the solar wind plasma continually expose the Earth's atmosphere and its magnetic fields. The solar wind, flares, and coronal mass ejections
show the interactions with the Earth's atmosphere and magnetic fields (${\rm e.g.,}$ geomagnetic disturbances), and may affect the space weather, communications, navigation systems, and astronauts \citep{ref:1929MNRAS..89..456C, parker1958interaction,ref:1968ApJ...151.1155H, ref:1971JGR....76.5202F, ref:1978GeoJ...54..547P, ref:1982JGR....87.9077Y, ref:1987JGR....92.5896C, ref:1991JGR....96.7831G, ref:2003JGRA..108.1246B,Wheatland2005, ref:2010RvGeo..48.4001G,chane2015simulations,Cranmer2017,Raboonik2017,farhang2018principle,nasibe}.

\cite{1965SSRv....4..666P} proposed the isothermal model for solar wind. In his model, the proton temperature anisotropy near the Earth plasma was not justified. After \cite{1965SSRv....4..666P}, several attempts have been made to investigate the behavior of the solar wind \citep{ref:1965JGR....70.4175W, ref:1971ApJ...166..669D, ref:1971ApJ...170..319D, ref:1972RSPSA.328..185R}.
\cite{meyer2007basics} studied the solar wind from various perspectives, and \cite{ref:2006LRSP....3....1M} considered wave-particle interactions in solar wind dynamics.

Observations revealed low values for the density of the solar wind. This wind mostly
originates from the polar coronal holes during the solar minimum \citep{ref:1995SSRv...72...49G, ref:2000JGR...10510419M, ref:2008GeoRL..3518103M}. The particle distributions for fast solar wind deviate from the Maxwellian distribution \citep{ref:1980SoPh...67..393L}. Also, the solar wind plasma can be considered collisionless. \cite{ref:1983bpp..conf....1K} presented a formulation for the collisional and collisionless plasma, which is useful for studying the solar wind.

In the kinetic study of the fast solar wind with non-Maxwellian distributions, the different sets of coupled equations partly agree with the solar wind observational data \citep[e.g.,][]{demars1990solar, ref:1991P&SS...39..435D, lie200116, ref:2011ApJ...743..197C}.
\cite{ref:1997PhPl....4.3974S} developed a set of fluid momentum equations that describe the kinetic Landau damping for the plasma. They also considered the Coulomb collisions for the particles. \\
The particle distribution function is a key aspect of the study of the plasma wave-particle interactions and instabilities.
For a homogeneous and isotropic plasma, the Maxwellian distribution determines the macroscopic parameters of the plasma in the thermal equilibrium and collisional condition \citep{ref:2004fopp.book.....B},
\begin{eqnarray}\label{maxwell}
f_{M}=\frac{n m^{1/2}}{(2 \pi k_{B} T)^{1/2}}\exp\left(-\frac{m v^{2}}{2 k_{B}T}\right),
\end{eqnarray}
where $n, m, k_{B}, T,$ and $v$ represent the number density, particle mass, Boltzmann constant, temperature, and velocity, respectively.
There are usually non-equilibrium conditions in the geophysical and space plasma, as the collisionless systems and the distributions of some high-energy particles deviate from the Maxwellian \citep{livadiotis2017kappa}.
With this objective in mind, the Kappa distribution function was proposed \citep[e.g.,][]{ref:1968ASSL...10..641O, ref:JGR:JGR7141,ref:2001Ap&SS.277..195P}:
\begin{eqnarray}
&&f_{\kappa}=\frac{n}{(\pi \theta ^{2}_{\kappa})^{3/2}} \frac{\Gamma(\kappa +1)}{\Gamma (\kappa -\frac{1}{2})}\left(1+\frac{v^{2}}{\kappa \theta ^{2}_{\kappa}}\right)^{-(\kappa+1)},\\
&& \theta^{2}_{\kappa}=\left(1-\frac{3}{2\kappa}\right)\left(\frac{2k_{B}T}{m}\right),\nonumber
\end{eqnarray}
where ${\kappa}$ is an index representing a deviation from the Maxwellian distribution and $\Gamma$ indicates the gamma function. Within the limit of large $\kappa$ $(\kappa \to \infty)$, the Kappa distribution tends to the Maxwellian one \citep{ref:2010SoPh..267..153P, livadiotis2013understanding}.

The Bi-Maxwellian distribution could explain the temperature anisotropy for the solar wind protons \citep{demars1990solar, ref:1991P&SS...39..435D, ref:2011ApJ...743..197C}.
The tail (particles with high speeds and energies) of the electron distribution is well described by the Kappa or power-law distributions \citep{zouganelis2004transonic}. The electron distribution can be classified into two categories: thermal core and suprathermal halo population \citep{ref:JGR:JGR7141, ref:2001Ap&SS.277..195P}.

Observations showed that the wave turbulence has a significant effect on the propagation of the solar wind \citep{ref:1968ApJ...153..371C}. This could be responsible for the heating and acceleration of the solar wind.

\cite{ref:2015NatCo...6E7813M} verified the existence of the Alfv\'en wave in the coronal open magnetic field regions as one of the reasons for the acceleration of the solar wind. The $Voyager's$ observations gave evidence for the presence of the Alfv\'en wave fluctuations in the solar wind up to $8$ AU from the Sun \citep{ref:1987JGR....9211021R,ref:2005LRSP....2....4B}.
Landau damping is an important mechanism for the Alfv\'en wave damping in collisionless plasma \citep{ref:1996JGR...101.5085L,ref:2013ApJ...774..139T}. In this mechanism, the oscillatory modes for plasma damp in the collisionless regime of a plasma. In non-Maxwellian distributions with suprathermal particles, the probability of  Landau damping has a high value \citep{ref:2009PhPl...16e2106B, ref:2010SoPh..267..153P, ref:2011PhPl...18a2307R, ref:1742-6596-516-1-012013}.
\cite{ref:0295-5075-113-2-25001} suggested the heating of particles in the inhomogeneous plasma related to the kinetic Alfv\'en wave (KAW) Landau damping.

The existence of anisotropic temperatures in plasma is a major reason for the application of the non-Maxwellian distribution for the particles.
The appearance and growth of instabilities are the results of deviation from the isotropy temperature \citep{ref:2017Ap&SS.362...13S}.

\cite{ref:2006GeoRL..33.9101H} studied the oblique, mirror, and oblique firehose instabilities using the WIND /SWE observational values and the temperature ratio (the ratio of the perpendicular component of temperature to the parallel component). \cite{ref:2006JGRA..111.3105K} showed that the mirror and cyclotron instabilities control the anisotropy for $T_{\bot p}/T_{\| p}>1$, and that the firehose instability controls the anisotropy for $T_{\bot p}/T_{\| p}<1$. The several instability mechanisms and wave turbulences in the solar corona and solar wind have been widely investigated \citep[e.g.,][]{Chandran2018,2018ApJ...853..190S,2018ApJ...860...17S}. 

\cite{ref:2011ApJ...743..197C} proposed a 1D, two-fluid model for the solar wind. They considered Maxwellian and Bi-Maxwellian distribution functions for the electrons and protons, respectively. They derived a set of coupled equations for the protons in parallel and perpendicular to the ambient magnetic field direction. The equations for quantities related to the system of electrons are also coupled with equations corresponding to the system of protons. They considered the low-frequency Alfv\'en wave in the wave-particle interactions and calculated heating rates of protons in two directions and the total heating rate of electrons. By applying the moments of the Vlasov equation, they derived a set of eight coupled equations. The set of equations in the solar wind conditions was solved with the \cite{ref:1997JGR...10214661H} method. 

In this paper, we extend the \cite{ref:2011ApJ...743..197C} model for the fast solar wind in the framework constructed by \cite{ref:1997PhPl....4.3974S} by applying the Kappa-Maxwellian distribution for electrons instead of the Maxwellian distribution. Consequently, we obtain separate equations for the components of electron temperature and heat flux. Using the zeroth- to fourth-order moments of the Vlasov equation, we derive a set of eleven coupled equations (instead of the 8 coupled equations given by \cite{ref:2011ApJ...743..197C}).
 
We solved the equations by applying the Iterated Crank-Nicolson (ICN) numerical method.
Discretizing equations with the ICN method has second-order accuracy in space and time, which offers accurate computational results.

The details of the derivation of the set of the coupled equations for the present two-fluid model are given in Section \ref{sec:equ}. The instabilities driven by temperature anisotropy are presented in Section \ref{instability}. Calculations of the heating rate for electrons and protons are provided in Section \ref{heat}. In Section \ref{num}, we present a numerical method for solving the 11 coupled equations. Numerical results are presented in Section \ref{res}, and they are compared with observations and previous studies. A conclusion is given in Section \ref{conclu}.
\section{Equations of the two-fluid solar wind model}\label{sec:equ}
\begin{figure}[h!]
	\centering
	\includegraphics[width=18cm]{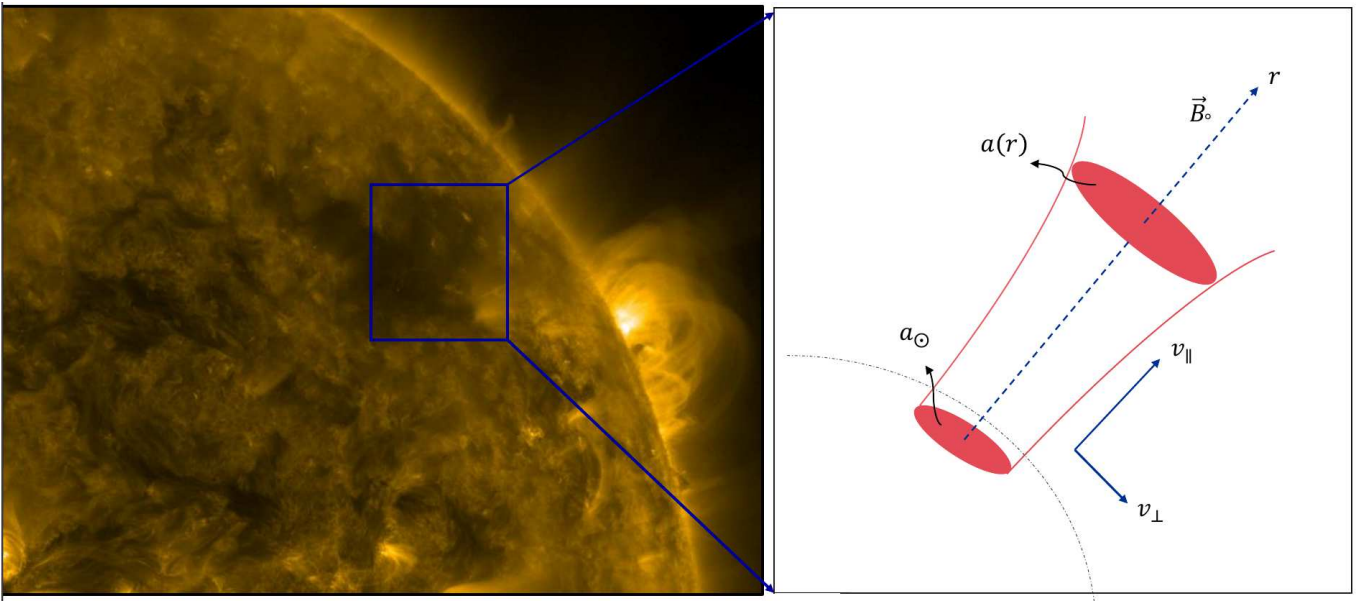}
	\caption{Left: an image at 171 $\mathring{A}$ observed by the $Solar Dynamic ~Observatory (SDO)$/Atmospheric Imaging Assembly (AIA) on 2013:07:07.
		Right: a schematic representation of an open magnetic flux tube on the Sun. The parameters $a_{\odot}$ and $a(r)$ are the cross section of the flux tube on the photosphere and at a distance $r$ from the Sun, respectively.}\label{fig1}
\end{figure}

The formulation of the present model for the
collisionless magnetohydrodynamics (MHD) is based on \cite{ref:1983bpp..conf....1K}. For this purpose, a thin open magnetic flux tube 
originating from a solar coronal hole along with the solar radii is considered.
A cylindrical coordinate with the z-axis along the magnetic field is used (Figure \ref{fig1}). The Sun's rotation is not considered in the calculations.

The fundamental variables are as follows: the mass density $\rho$, the fluid velocity
${\bf U}={\bf v}_{E}+u_{\|}{\bf \hat{b}}$ (which is the same for
electrons and protons), the magnetic field
${\bf B}$, the proton distribution function $f_{p}$, the electron distribution function $f_{e}$,
and the parallel component of the electric field $E_{\|}={\bf \hat{b}}.{\bf E}$ (${\bf \hat{b}}={\bf B}/B$ is a unit vector along the magnetic field). After that, we used the notation used by \cite{ref:2011ApJ...743..197C}.

In Kulsrud's formulation, the Vlasov equation is given by \citep{ref:1983bpp..conf....1K,ref:1997PhPl....4.3974S}
\begin{eqnarray}\label{eq1}
\frac{\partial}{\partial t}(f_{s}B)+ \nabla.[ f_{s}B(v_{\|}{\bf \hat{b}}+{\bf v}_{E})]+ \frac{\partial}{\partial v_{\|}}
\times [f_{s}B(-{\bf \hat{b}}.\frac{D{\bf v}_{E}}{Dt}-\mu{\bf \hat{b}}.\nabla B+\frac{e_{s}E_{\|}}{m_{s}})]=0,
\end{eqnarray}
where $s$ indicates particle species ($p$ for proton and $e$ for electron), $f_{s}$ is the
particle distribution function, $m_{s}$ and $e_{s}$ are
the mass and charge, and ${\bf v}$ is the velocity of the particle
($v_{\|}={\bf \hat{b}}.{\bf v},~{\bf v}_{E}=c({\bf E}\times {\bf B})/B^{2},~\mu=v_{\bot}^2 /2B$).

The total derivative is defined by
$D/Dt=\partial/\partial t+(v_{\|}{\bf \hat{b}}+{\bf v}_{E}).{\bf \nabla}$. The distribution function
($f_{s}$) is a function of position ${\bf r}$~(heliocentric distance in the solar wind model), time $t$, magnetic moment $\mu$, and the parallel component of velocity $v_{\|}$.

The collisionless MHD equations can be derived by evaluating different orders of the velocity moments (Equation \ref{eq1}) and the modified electromagnetic Maxwell equations.
Given the limit of low Alfv$\acute{e}$n speed ( $v_{A}^{2}\leq c^{2}$),
the continuity and momentum equations are given by \citep{ref:1997PhPl....4.3974S},
\begin{eqnarray}
&&\frac{\partial \rho}{\partial t}+ \nabla(\rho {\bf U})=0,
\label{eq2}\\
&&\rho\left(\frac{\partial {\bf U}}{\partial t}+{\bf U}.\nabla{\bf U}\right)= \frac{(\nabla \times {\bf B})\times {\bf B}}{4\pi}-
\nabla.{\bf P}
-\frac{GM_{\bigodot}\rho}{r^{2}}{\bf \hat b}-\frac{1}{2}\frac{\partial E_{W}}{\partial r}{\bf \hat b}.
\label{eq3}
\end{eqnarray}
The third and fourth terms on the right side of Equation (\ref{eq3}) are the gravitational acceleration and the Alfv\'en wave pressure force, respectively.
The pressure tensor ${\bf P}$ can be written as \citep{goedbloed2004principles},
\begin{eqnarray}\label{eq5}
{\bf P}=\sum_{s} p_{\bot s}({\bf I}-{\bf \hat{b}\hat{b}})+\sum_{s} p_{\|s} {\bf \hat{b} \hat{b}},
\end{eqnarray}
where ${\bf I}$ is an unit dyadic. The parallel and perpendicular components of the pressure tensor are given by
\begin{eqnarray}&&\label{eq6}
p_{\bot s}:= \frac {m_{s}}{2}\int f_{s}v_{\bot}^{2} d^{3}v,
\\&&\label{eq7}
p_{\|s} := m_{s} \int f_{s} (v_{\|}-U_{\|})^{2} d^{3}v.
\end{eqnarray}
The number density is defined as
\begin{eqnarray}\label{eq9}
n_{s}:=\int f_{s}d^{3}v.
\end{eqnarray}
The induction equation is introduced by
\begin{eqnarray}\label{eq4}
\frac{\partial{\bf B}}{\partial t}=\nabla\times({\bf U}\times{\bf B}).
\end{eqnarray}
In the lowest order in $1/e$, the electrostatic Poisson equation for charges and number densities is reduced to the condition, $\sum_{s} e_{s} n_{s}=0$ \citep{ref:1983bpp..conf....1K}. Furthermore, we assume $n=n_{p}=n_{e}$. The electron contribution to mass density is not considered, and the total mass density is $\rho=nm_{p}$. 
The perpendicular pressure $p_{\bot s}$ satisfies \citep{ref:1997PhPl....4.3974S,ref:2006ApJ...637..952S,ref:2011ApJ...743..197C},
\begin{eqnarray}\label{eq10}
\rho B \frac{d}{dt}(\frac{p_{\bot s}}{\rho B})= -{\bf \nabla}.(q_{\bot s}{\bf \hat{b}})- q_{\bot s}{\bf \nabla}.{\bf \hat{b}}+ \frac{\nu_{s}}{3}(p_{\| s}-p_{\bot s}){\bf ,}
\end{eqnarray}
where $\nu_{s}$ is the Coulomb collision frequency for the energy exchange between particles.
Moreover, the parallel component of pressure $p_{\| s}$ obeys
\begin{eqnarray}\label{eq11}
\frac{\rho^{3}}{2B^{2}}\frac{d}{dt}(\frac{B^{2}p_{\| s}}{\rho ^{3}})= - {\bf \nabla}.(q_{\| s}{\bf \hat{b}})+q_{\bot s} {\bf \nabla.\hat{b}}+\frac{\nu_{s}}{3}(p_{\bot s}-p_{\| s}),
\end{eqnarray}
where the perpendicular and parallel components of the heat flux are defined by
\begin{eqnarray}
q_{\bot s}&:=&m_{s} \int f_{s}\mu B (v_{\|}-U_{\|}) d^{3}v,\label{eq12}\\
q_{\|s}&:=& \frac{m_{s}}{2}\int f_{s}(v_{\|}-U_{\|})^{3} d^{3}v.\label{eq13}
\end{eqnarray}
The perpendicular heat flux $q_{\bot s}$ is given by
\begin{eqnarray}\label{eq16}
\rho ^{2} \frac{d}{dt}(\frac{q_{\bot s}}{\rho ^{2}})+ \nu_{s} q_{\bot s}&=&- {\bf \nabla}.(r_{\|\bot}{\bf \hat{b}})+ \frac{p_{\bot s}}{\rho} {\bf \hat{b}}. \nabla p_{\| s}\\ \nonumber
&+& [\frac{p_{\bot s}(p_{\| s}-p_{\bot s})}{\rho}+r_{\bot\bot}-r_{\|\bot}] {\bf \nabla . \hat{b}},
\end{eqnarray}
and for $q_{\| s}$ is
\begin{eqnarray}\label{eq17}
\frac{\rho ^{4}}{B^{3}}\frac{d}{dt}(\frac{B^{3}q_{\| s}}{\rho ^{4}})+ \nu_{s} q_{\| s}&=&-\frac{1}{2} {\bf \nabla}.(r_{\|\|}{\bf \hat{b}}) + \frac{3p_{\| s}}{2\rho} {\bf \hat{b}}. \nabla p_{\| s} \\ \nonumber
&+& \frac{3}{2}[\frac{p_{\| s}(p_{\| s}-p_{\bot s})}{\rho}+r_{\bot\|}] {\bf \nabla . \hat{b}}.
\end{eqnarray}
The fourth-order moments of the Vlasov equation ($r_{\bot\bot},r_{\| \bot},r_{\|\|}$) are introduced by
\begin{eqnarray}&&\label{eq18}
r_{\bot\bot}:=m_{s}\int f_{s} \mu^{2} B^{2} d^{3}v,
\\&&\label{eq19}
r_{\| \bot}:=m_{s} \int f_{s} \mu B (v_{\|}-U_{\|})^{2} d^{3} v,
\\&&\label{eq20}
r_{\|\|}:= m_{s} \int f_{s} (v_{\|}-U_{\|})^{4} d^{3} v.
\end{eqnarray}
We consider the Kappa-Maxwellian distribution function for the electrons as follows:
\begin{eqnarray}\label{eq21}
f_{e}=f_{\kappa M}(v_{\|},v_{\bot})=\frac{n}{\pi^{3/2}\theta_{\bot e}^{2}\theta_{\| e}} \frac{\Gamma(\kappa+1)}{\kappa^{3/2}\Gamma(\kappa-\frac{1}{2})}
 \left(1+\frac{(v_{\| e}-U_{\| e})^{2}}{\kappa\theta_{\| e}^{2}}\right)^{-\kappa} \exp\left(-\frac{v_{\bot e}^{2}}{\theta_{\bot e}^{2}}\right),
\end{eqnarray}
where the parallel and perpendicular components of the thermal velocities are defined as
\[\theta_{\| e}=\left(\frac{2\kappa-3}{\kappa} \right)^{\frac{1}{2}}\left( \frac{k_{B}T_{\| e}}{m_{e}} \right)^{\frac{1}{2}} , \qquad  \theta_{\bot e}=\left( \frac{2 k_{B}T_{\bot e}}{m_{e}} \right)^{\frac{1}{2}}.\]
For the Kappa distribution, the spectral index $\kappa$ is a free parameter and varies from $1.5$ to infinity \citep{ref:2010SoPh..267..153P}.
The Bi-Maxwellian distribution function for protons is introduced by
\begin{eqnarray}
f_{p}=f_{BM}=\frac{n m_{p}^{3/2}}{(2\pi k_{B})^{3/2}T_{\bot p}T_{\| p}^{1/2}}
\exp\left(-\frac{m_{p} \mu_{p} B}{k_{B}T_{\bot p}}-\frac{m_{p}(v_{\| p}-U_{\| p})^{2}}{2k_{B}T_{\| p}}\right).
\end{eqnarray}
In the reminder of this section, we explore the explicit effects of the electron and proton distribution functions on the quantities of the system of the two-fluid model. The relation between the fourth-order moments, $r_{\bot\bot}$, $r_{\bot\|}$, $r_{\|\|}$
(Equations \ref{eq18}, \ref{eq19}, and \ref{eq20}) and main quantities ($n$, $p$, $T$, $U$, etc.), are derived.\\
Suppose a straight flux tube with a magnetic field along the solar radius $r$. Then we have 
\begin{eqnarray}\label{eq40}
{\bf \hat{b}.\nabla} \equiv \frac{\partial}{\partial r},~~~~~~~~~~~{\bf \nabla . \hat{b}}=\frac{1}{a}\frac{\partial a}{\partial r}, 
\end{eqnarray}
where $a$ is the cross section of the flux tube  \citep{ref:1976SoPh...49...43K}. Additionally, we assume all the
quantities are the function of solar radii ($r$) (along with the axis of the flux tube)
and are considered as axially symmetric (independent of $\phi$ in cylindrical coordinate).

The continuity, momentum, and pressure equations, which are the same for both electrons and protons, are derived from the zeroth- to second-order moments of the Vlasov equation. Owing to the different nature of the distributions for electrons and protons, the equations for the electron heat flux are different from the proton heat flux.

The set of variables depending on the time ($t$) and radial coordinate ($r$) comprises the following: number density
$n$ (proton or electron), outflow velocity $U$ (proton or electron),
perpendicular and parallel electron temperatures $T_{\bot e}$
and $T_{\| e}$, perpendicular and parallel proton temperatures
$T_{\bot p}$ and $T_{\| p}$, electron heat fluxes $q_{\bot e}$
and $q_{\| e}$, proton heat fluxes $q_{\bot p}$ and $q_{\| p}$, and wave energy $E_{W}$.

Using Equation (\ref{eq40}), the continuity equation (Equation \ref{eq2}) gives
\begin{eqnarray}\label{eq23}
\frac{dn}{dt}=-\frac{n}{a}\frac{\partial}{\partial r}(aU),~~~~~~~~~~~~~~~n=n_{p}=n_{e}.
\end{eqnarray}
Substituting the pressure tensor from Equation (\ref{eq5}) with the momentum equation (Equation \ref{eq3}), and after some algebra manipulation, one finds
\begin{eqnarray}\label{eq24}
&&\frac{dU}{dt}=-\frac{k_{B}}{\rho}\frac{\partial}{\partial r}\left(n(T_{\|e}+T_{\|p})\right)+\frac{k_{B}[(T_{\bot p}-T_{\| p})+(T_{\bot e}-T_{\| e})]}{m_{p}a}\frac{\partial a}{\partial r} - \frac{GM_{\bigodot}}{r^{2}}-\frac{1}{2\rho}\frac{\partial E_{W}}{\partial r}.
\end{eqnarray}
By calculating the fourth-order moments (Equations \ref{eq18} - \ref{eq20})) according to the related distribution function, we find parallel and perpendicular electron heat fluxes:
\begin{eqnarray}\label{eq25}
\rho ^{2} \frac{d}{dt}(\frac{q_{\bot e}}{\rho ^{2}})+ \nu_{e} q_{\bot e}=
-\frac{nk_{B}^{2}T_{\|e}}{m_{e}}\frac{\partial T_{\bot e}}{\partial r}
+\frac{nk_{B}^{2}T_{\bot e}(T_{\bot e}-T_{\|e})}{m_{e} a}\frac{\partial a}{\partial r},
\end{eqnarray}
\begin{eqnarray}\label{qpare}
\frac{\rho ^{4}}{B^{3}}\frac{d}{dt}\left(\frac{B^{3}q_{\| e}}{\rho ^{4}}\right)+ \nu_{e} q_{\| e}&=&
+\frac{3n^{2}k_{B}^{2}T_{\| e}^{2}}{(5-2\kappa)\rho}\frac{1}{a}\frac{\partial a}{\partial r}\nonumber\\
+\frac{3(2\kappa-1)}{2(5-2\kappa)}\frac{n^{2}k_{B}^{2}T_{\| e}}{\rho}\frac{\partial T_{\| e} }{\partial r}& +& \frac{3}{(5-2\kappa)}\frac{n k_{B}^{2}T_{\| e}^{2}}{\rho}\frac{\partial n}{\partial r}.
\end{eqnarray}
The proton heat flux equations are given by \citep[e.g.,][]{ref:2011ApJ...743..197C}
\begin{eqnarray}\label{eq27}
\rho ^{2} \frac{d}{dt}(\frac{q_{\bot p}}{\rho ^{2}})+ \nu_{p} q_{\bot p}= -\frac{nk_{B}^{2} T_{\| p}}{m_{p}}\frac{\partial T_{\bot p}}{\partial r}
+ \frac{nk_{B}^{2}T_{\bot p}(T_{\bot p}-T_{\| p})}{m_{p} a} \frac{\partial a}{\partial r},
\end{eqnarray}
\begin{eqnarray}\label{eq28}
\frac{n^{4}}{B^{3}} \frac{d}{dt} \left(\frac{B^{3} q_{\| p}}{n^{4}} \right)+\nu_{p}q_{\| p}=-\frac{3nk_{B}^{2}T_{\| p}}{2m_{p}}\frac{\partial T_{\| p}}{\partial r}.
\end{eqnarray}
Now, we use the temperatures instead of the pressures in the equations for both electrons and protons.
This can be done by substituting $p_{\bot}=nk_{B}T_{\bot}$ and $p_{\|}=nk_{B}T_{\|}$ in Equations (\ref{eq10}) and (\ref{eq11}) for electron temperatures,
\begin{eqnarray}\label{Tper}
B n k_{B}\frac{d}{dt}(\frac{T_{\bot e}}{B})&=&Q_{\bot e}-\frac{1}{a^{2}}\frac{\partial}{\partial r}(a^{2}q_{\bot e})+\frac{1}{3}\nu _{e}nk_{B}(T_{\| e}-T_{\bot e})\nonumber \\&+&2\nu_{ep} n k_{B}(T_{p}-T_{\bot e}),
\end{eqnarray}
\begin{eqnarray}\label{Tpar}
\frac{n^{3}k_{B}}{2B^{2}}\frac{d}{dt}(\frac{B^{2}T_{\|e}}{n^{2}})=Q_{\| e}-\frac{1}{a}\frac{\partial}{\partial r}(a q_{\| e})+ \frac{q_{\bot e}}{a} \frac{\partial a}{\partial r}\nonumber \\+\frac{1}{3} \nu _{e} n k_{B}(T_{\bot e}-T_{\| e})+\nu _{ep}nk_{B}(T_{p}-T_{\| e}).
\end{eqnarray}
The parallel and perpendicular components of the proton temperature obey the following equations:
\begin{eqnarray}\label{Tparp}
B n k_{B}\frac{d}{dt}(\frac{T_{\bot p}}{B})&=&Q_{\bot p}-\frac{1}{a^{2}}\frac{\partial}{\partial r}(a^{2}q_{\bot p})+\frac{1}{3}\nu _{p}nk_{B}(T_{\| p}-T_{\bot p})\nonumber \\&+&2\nu_{pe} n k_{B}(T_{e}-T_{\bot p}),
\end{eqnarray}
\begin{eqnarray}\label{Tperp}
\frac{n^{3}k_{B}}{2B^{2}}\frac{d}{dt}(\frac{B^{2}T_{\|p}}{n^{2}})=Q_{\| p}-\frac{1}{a}\frac{\partial}{\partial r}(a q_{\| p})+ \frac{q_{\bot p}}{a} \frac{\partial a}{\partial r}\nonumber \\+\frac{1}{3} \nu _{p} n k_{B}(T_{\bot p}-T_{\| p}) +\nu _{pe}nk_{B}(T_{e}-T_{\| p}).
\end{eqnarray}
In Equations (\ref{Tper})-(\ref{Tperp}), the quantities ($Q_{\bot e}$, $Q_{\| e}$) and ($Q_{\bot p}$, $Q_{\| p}$) are the heating rates per unit volume for electrons and protons, respectively. In Section \ref{heat}, the details of heating rates are given.

Finally, the last equation for wave energy $E_{W}$ is given by \citep{ref:1970PhFl...13.2710D}
\begin{eqnarray}\label{eq33}
\frac{\partial E_{W}}{\partial t}+\frac{1}{a}\frac{\partial}{\partial r}[a(U+v_{A})E_{W}]+\frac{E_{W}}{2a}\frac{\partial}{\partial r}(aU)=-Q,
\end{eqnarray}
where the Alfv\'en speed is $v_{A}=\frac{B}{\sqrt{4\pi \rho}}$, the total heating rate is 
$Q=Q_{\bot e}+Q_{\|e}+Q_{\bot p}+Q_{\| p}$, and the total temperature is defined as
$T=\frac{T_{\|}+2T_{\bot}}{3}$.
Using Equations (\ref{eq24}) and (\ref{Tper})- (\ref{eq33}), the total energy equation is obtained:
\begin{eqnarray}
\frac{\partial E_{\rm tot}}{\partial t}+\frac{1}{a}\frac{\partial}{\partial r}(aF_{\rm tot})=0,
\end{eqnarray}
where $E_{\rm tot}$ is the total energy density and is defined as
\begin{eqnarray}\label{energydensity}
E_{\rm tot}=\frac{\rho U^{2}}{2}-\frac{GM_{\odot}\rho}{r}+nk_{B}(T_{\perp e}+\frac{T_{\|e}}{2}+T_{\perp  p}+\frac{T_{\|p}}{2})+E_{W},
\end{eqnarray}
and $F_{\rm tot}$ is the total energy flux and is defined as
\begin{eqnarray}\label{energyflux}
F_{\rm tot}=\frac{\rho U^{3}}{2}&-&\frac{UGM_{\odot}\rho}{r}+Unk_{B}(T_{\perp e}+\frac{3T_{\|e}}{2}+T_{\perp p}+\frac{3T_{\|p}}{2})\nonumber\\
&+&q_{\rm total}+(\frac{3U}{2}+v_{A})E_{W},
\end{eqnarray}
where 
$q_{\rm tot}=q_{\perp e}+q_{\perp p}+q_{\| e}+q_{\| p}$.
In Equation (\ref{energyflux}), the first, second, third, and last terms are the kinetic energy flux, gravitational potential energy, enthalpy flux, and Alfv\'en wave enthalpy flux, respectively.\\
In the steady state, $\frac{\partial E_{\rm tot}}{\partial t}=0$, the total energy flux is conserved and constant along the flux tube (solar radius).

\subsection{The Limit of Large $\kappa$}
Generally, observations of the space plasma showed that the tail of the distribution is likely to be the power-law function. It is important to use the Kappa distribution as a non-Maxwellian distribution for such plasmas. The Kappa distribution approach to the Maxwellian distribution for large $\kappa$ index ($\kappa$ tends to $\infty$) and for finite $\kappa$ differs from the Maxwellian.
Therefore, we expect that, in the limit of large $\kappa$, the set of equations derived in the presence of the  Kappa-Maxwellian approach to the Bi-Maxwellian set.

The parallel component of the electron heat flux (Equation \ref{qpare})
with the Kappa-Maxwellian distribution has an explicit dependency on the $\kappa$ index.
In the limit of the large $\kappa$, and by setting $\rho=m_{p}n$, the equation for the parallel  electron heat flux Equation (\ref{qpare}) can be rewritten as 
\begin{eqnarray}\label{new}
\frac{n^{4}}{B^{3}}\frac{d}{dt}\left( \frac{B^{3} q_{\|e}}{n^{4}}\right)+\nu_{e}q_{\|e}=-\frac{3n^{2}k_{B}^{2}T_{\|e}}{2\rho}\frac{\partial T_{\|e}}{\partial r},
\end{eqnarray}
Equation (\ref{new}) has the same form for the parallel heat flux obtained for the Bi-Maxwellian distribution function. Other quantities are coupled with the electron heat flux and depend on the $\kappa$ index.
\section{Instabilities driven by temperature anisotropy}\label{instability}
Anisotropic behavior of the temperature of particles (electron and proton) leads to plasma instabilities \citep{gary1996whistler,ref:2017Ap&SS.362...13S}. The proton and electron temperature anisotropy ratios are defined by 
$R_{p}=T_{\perp p}/T_{\parallel p}$ and $R_{e}=T_{\perp e}/T_{\parallel e}$, respectively. Observations show that for plasma stability, these ratios should remain in the specific ranges. The oblique firehose and mirror instabilities restrict the $R_{p}$ for the lower and upper limits for protons. Also, mirror and Whistler instabilities control the lower and upper limits of the $R_{e}$ for electrons \citep{kalman1968anisotropic,gary1996whistler,kasper2002wind,gary2006linear,ref:2006GeoRL..33.9101H,bale2009magnetic}. 
The values of both $R_{p}$ and $R_{e}$ are related to the plasma beta parameter $(\beta_{\parallel})$. In the case of  $\gamma_{max}\leqslant
10^{-3} \Omega_{p}$, where $\gamma_{max}$ is the maximum growth rate of instabilities and $\Omega_{p}$ is the proton cyclotron frequency, the relations for instabilities (mirror and oblique firehose) of protons temperature anisotropy are given by
\begin{eqnarray}
R_{p,m}=1+0.77(\beta_{\parallel p}+0.016)^{-0.76}, &~~~~~~~~~~ R_{p,f}=1-1.4(\beta_{\parallel p}+0.11)^{-1},
\end{eqnarray}
in which $\beta_{\parallel p}= \frac{8\pi n k_{B} T_{\parallel p}}{B^{2}}$ \citep{ref:2006GeoRL..33.9101H}. The instabilities (Whistler and mirror) for electrons temperature anisotropy are given by 
\begin{eqnarray}
R_{e,w}=1+0.15\beta_{\parallel e}^{-0.56}, &~~~~~~~~~~ R_{e,m}=1+0.53\beta_{\parallel e}^{-0.64},
\end{eqnarray}
where $\beta_{\parallel e}= \frac{8\pi n k_{B} T_{\parallel e}}{B^{2}}$ \citep{gary2006linear}.

Following \cite{ref:2011ApJ...743..197C}, we insert the temperature-driven anisotropy effects in terms of $\nu_{\rm inst}$ for protons as follows:
\begin{eqnarray}
\nu_{p,\rm inst}=\nu_{0}\exp \left(\frac{12(R_{p}-R_{p,m})}{R_{p,m}}\right)+\nu_{0}\exp \left(\frac{12(\bar{R}_{p,f}-R_{p})}{\bar{R_{p,f}}}\right),~~~~~~\bar{R}_{p,f}=\max(R_{p,f},10^{-6}),
\end{eqnarray}
and $\nu_{e,\rm inst}$ for electrons is as follows:
\begin{eqnarray}
\nu_{e,\rm inst}=
\nu_{0}\exp\left(\frac{12(R_{e}-R_{e,w})}{R_{e,w}}\right)+\nu_{0}\exp\left(\frac{12(R_{e,m}-R_{e})}{R_{e,m}}\right),
\end{eqnarray}
where $\nu_{0}=0.02\sqrt{GM_{\odot}/R_{\odot}^{3}}$. Finally, $\nu_{p}$ and $\nu_{e}$ are defined as 
\begin{eqnarray}
\nu_{p}=\nu_{pp}+\nu_{p,\rm inst}~, &~~~~~~~~~~\nu_{e}=\nu_{ee}+\nu_{e,\rm inst},
\end{eqnarray}
in which $\nu_{pp}$ and $\nu_{ee}$ are the proton-proton and electron-electron Coulomb collision frequency \citep{ref:1975P&SS...23..437S}.

\section{Heating rates}\label{heat}
The nature of the turbulence dissipation of the solar wind is not yet to be well understood, but observations and analytical calculations have verified the Alfv\'enic turbulence effects in the damping mechanisms  \citep[e.g,][]{jiang2009cascade,chen2010anisotropy,vranjes2010kinetic,ref:2012ApJ...745L...9S, ref:2013ApJ...774..139T,2015MNRAS.454.3697M, ref:2016JGRA..121.7349W, schreiner2017model}.

For the plasma with spatial scales much larger than the particle's mean free path, the MHD approach provides a suitable description of the propagation without damping the fast, intermediate (Alfv\'en), and slow modes. Hence, undamped plasma waves cascade to the small scales.

 In general, an Alfv\'en wave is a type of MHD waves in which the ions vibrate due to the disturbing of the magnetic field lines in a magnetized plasma. Both transverse and longitudinal Alfv\'en waves have been detected \citep[e.g.,][]{hollweg1981alfven, amagishi1986experimental}. In transverse Alfv\'en waves (shear Alfv\'en wave) both the disturbance of the magnetic field and the motion of ions are in the same direction and perpendicular to the direction of the wave vector (propagation direction) \citep{alfven1947granulation, cramer2011physics, priest2012solar, esmaeili2017behavior}.  

The Alfv\'en modes remain undamped until the structures of the plasma reach the size of the proton gyroradius. The KAW fluctuations appear in the turbulence cascades of Alfv\'en waves and move their fluid scale to the smaller structures (kinetic scale) \citep{zhao2011kinetic,gershman2017wave}, thus creating non-thermal particles. For more details, see
\cite{howes2008inertial,ref:2015ASSL..407..123H}, \cite{howes2011gyrokinetic}.

Depending on the dissipation mechanism, the total heating rate $Q$ is divided between the electrons ($Q_{\bot e},Q_{\| e}$) and protons ($Q_{\bot p}$, $Q_{\| p}$).
The contribution of each species to the total heating rate could be calculated by a numerical solution of the dispersion relation for the Alfv\'en wave.

By linearizing the Maxwell equations, the following dispersion relation is given in the Fourier space ($\omega , k$) \citep[e,g.,][]{ref:1998ApJ...500..978Q, ref:1992wapl.book.....S} as,
\begin{eqnarray}\label{eq39}
\bf{k}\times (\bf{k} \times \bf{E})+\frac{\omega^{2}}{c^{2}}\epsilon . \bf{E}=0,
\end{eqnarray}
where $E$, $\epsilon$,  $k$, $\omega$, and c represent the electric field perturbation, dielectric tensor, wave vector, wave frequency, and light speed, respectively. The dielectric tensor is related to the susceptibility tensor by 
\begin{eqnarray}\label{ep}
\epsilon_{ij}=\delta_{ij}+\sum_{s}\chi_{ij}^{s}.
\end{eqnarray}
The components of the susceptibility tensor are computed by \cite{ref:2007PhPl...14h2111C}.
To study the Alfv\'en wave and KAW interactions with plasma particles, we consider two ranges: $k_{\bot}\rho_{p}\sim1$ (for protons) and 
$k_{\bot}\rho_{p}\gg 1$ (for electrons).
The parallel wavenumber ($k_{\|}$) is obtained by the critical balance condition \citep{ref:1995ApJ...438..763G, ref:2000ApJ...539..273C, ref:2001ApJ...554.1175M, ref:2012PhPl...19e5901T}.
\cite{ref:1992wapl.book.....S} calculate particle damping rates. Following \cite{ref:2011ApJ...743..197C}, we calculate the parallel and perpendicular components of the electron and proton heating rates as 
\begin{eqnarray}
&&Q_{\|e}=\frac{(1+\gamma_{\|e}t_{c})Q}{1+\gamma_{\rm tot}t_{c}},\\
&&Q_{\bot e}=\frac{\gamma_{\bot e}t_{c}Q}{1+\gamma_{\rm tot}t_{c}},\\
&&Q_{\| p}=\frac{\gamma_{\| p}t_{c}Q}{1+\gamma_{\rm tot}t_{c}},\\
&&Q_{\bot p}=\frac{\gamma_{\bot p}t_{c}Q}{1+\gamma_{\rm tot}t_{c}},
\end{eqnarray}
where $t_{c}=\rho \delta v_{p}^{2}/Q$ is the time that the energy cascades at the scale $k_{\bot} \rho_{p}=1$. $\delta v_{p}$ is the root mean square (rms) of the Alfv\'en and/or KAW fluctuations. \\
Total heating rate per unit volume is introduced by \cite{ref:2011ApJ...743..197C} as
\begin{eqnarray}
Q=\frac{c_{d} \rho z^{-}_{\rm rms}(z_{\rm rms}^{+})^{2}}{4L_{\bot}},
\end{eqnarray}
where $c_{d}=0.75$ is a dimensionless number, and $z^{-}_{\rm rms}$ and $z_{\rm rms}^{+}$ are the Elsasser variables that satisfy the following equations:
\begin{eqnarray}
&&E_{W}=\frac{\rho (z_{\rm rms}^{+})^{2}}{4},\\
 &&z^{-}_{\rm rms}=\frac{L_{\bot}(U+v_{A})}{v_{A}}\arrowvert \frac{\partial v_{A}}{\partial r}\arrowvert.
\end{eqnarray}
$L_{\bot}$ is the correlation length scale due to the Alfv\'enic fluctuations.

\section{Numerical Method}\label{num}
In this study, we build a system of coupled nonlinear partial differential equations (Equations \ref{eq23}-\ref{eq33}) in the following general form:
\begin{eqnarray}\label{opt1}
\frac{\partial \psi(x,t)}{\partial t}= \mathcal{L}(\psi(x,t)),
\end{eqnarray}
where $\psi(x,t)$ and $\mathcal{L}$ are the vector of quantities and partial differential operator, respectively. We convert Equation (\ref{opt1}) to the Euler frame, and then discretize the equation(s) using the finite difference method in the spatial dimension by the mid-point approximation and forward in time as \citep{recktenwald2004finite, meis2012numerical, thomas2013numerical},
\begin{eqnarray}
\frac{\partial \psi(x,t)}{\partial x} &\approx& \frac{\psi_{i+1}^{j}-\psi_{i-1}^{j}}{x_{i+1}-x_{i-1}}\\ 
\frac{\partial \psi(x,t)}{\partial t} &\approx& \frac{\psi_{i}^{j+1}-\psi_{i}^{j}}{t_{j+1}-t_{j}}\nonumber
\end{eqnarray}
where the $i$ and $j$ indices represent the spatial and time steps, respectively. To solve the system of partial differential equations (Equations \ref{eq24}-\ref{eq33}), we implement the 
ICN method \citep{ref:2006PhRvD..73d4001L}, which is based on the prediction, correction, and averaging of the quantities.\\
The Crank-Nicolson method has second-order accuracy in both time and space \citep{ref:2000PhRvD..61h7501T}. In the method, two iterations are used to solve the Equation (\ref{opt1}). These two steps produce the iterative equations
\begin{eqnarray}
&&^{(1)}\tilde{\psi}^{n+1}_{i}= \psi_{i}^{n}+\Delta t  \mathcal{L}(\psi_{m}^{n}) ,\\
&&^{(1)}\bar {\psi}_{i}^{n+1/2} \equiv \frac{1}{2}\big(^{(1)}\tilde{\psi}^{n+1}_{i}+\psi_{i}^{n}\big),\\
&&^{(2)}\tilde{\psi}^{n+1}_{i}= \psi_{i}^{n}+\Delta t \mathcal{L}\big(^{(1)}\bar \psi_{m}^{n+1/2}\big) ,\\
&&^{(2)}\bar \psi_{i}^{n+1/2} \equiv \frac{1}{2}\big(^{(2)}\tilde{\psi}^{n+1}_{i}+\psi_{i}^{n}\big),\\
&&\psi^{n+1}_{i}= \psi_{i}^{n}+\Delta t \mathcal{L}\big(^{(2)}\bar \psi_{m}^{n+1/2}\big) ,
\end{eqnarray}
where $\tilde{\psi}$ and $\bar {\psi}$ are the predicted-corrected and averaged functions, respectively.
The value of the index $m$ depends on the order of operator $\mathcal{L}$ and second-order accuracy. For the first-order spatial derivative, we choose $m=i\pm1$. Each time step is adopted considering the stability condition of the ICN method.\\
To avoid the non-physical oscillations and also to increase stability in ICN outputs, we add the artificial diffusion term as $-D\frac{\partial^{2} \psi}{\partial x^{2}}$ to the right sides of equations, in which D is a positive constant.
 Empirically, we found that (a) the value of the diffusion constant need not all be equal for all equations; (b) for small values of the diffusion constants (0 $\leqslant$ D $\leqslant$ 5), the computational algorithm remains stable; (c) the diffusion terms for some quantities (e.g., fluid velocity, temperatures) are more important than those for others (e.g., number density).

We use a logarithmic grid in the space ($r$) with a growing size by increasing the distance from the origin due to rapid changes in the physical quantities near the Sun.
The parameter $r_{i}$ ($i=0,1,2,...,N+1$) extends from one solar radius (1 R$_{\odot}$) to one astronomical unit (1 AU). For a suitable computational time, we set the number of grids to $N=2000$. 
\subsection{Initial and Boundary Conditions}
We choose the following initial conditions (at $t = 0$) \citep{ref:2011ApJ...743..197C} for all grid points from $r_{0}$ to $r_{N+1}$ as

$~~~~~~~~~~~~~~~~~~~~~~~~~~~~~n=n_{\odot}U_{0}a_{\odot}/(Ua),~~~~~~~~U_{0}=U(r_{0}),$\\
$~~~~~~~~~~~~~~~~~~~~~~~~~~~~~~~~T_{\bot e} = T_{\| e} = T_{\bot p} = T_{\| p} = T_{\odot}(3-2R_{\odot}/r)(r/R_{\odot})^{-2/7},$\\
$~~~~~~~~~~~~~~~~~~~~~~~~~~~~~~~~U=(655~~ $km/s$)(1+20(R_{\odot}/r)^{3})^{-1},$\\
$~~~~~~~~~~~~~~~~~~~~~~~~~~~~~~~~E_{W}=n m_{p}(\delta v_{\odot})^{2},$\\
$~~~~~~~~~~~~~~~~~~~~~~~~~~~~~~~~q_{\bot p}= q_{\| p} = q_{\bot e}= q_{\| e} = 0.$

 We use the boundary conditions at $r_{0}$ close to the Sun \citep{ref:2011ApJ...743..197C}, 

 $~~~~~~~~~~~~~~~~~~~~~~~~~~~~~n=n_{\odot}=10^{8}~$ cm$^{-3}$,\\
 $~~~~~~~~~~~~~~~~~~~~~~~~~~~~~~~~T_{\|e} = T_{\bot e} = T_{\| p} = T_{\bot p} =T_{\odot}= 7 \times 10^{5} ~$K,\\
 $~~~~~~~~~~~~~~~~~~~~~~~~~~~~~~~~E_{W}=n_{\odot} m_{p}(\delta v_{\odot})^{2} , ~~~~~~~~~~~~~  \delta v_{\odot}=41.4~~$km/s.               

The rest value of the quantities $(q_{\|e}, q_{\bot e}, q_{\| p}, q_{\bot p},U)$ at $r_{0}$ is linearly extrapolated from their values in the next two grid points ($r_{1}$ and $r_{2}$).  Also, the open boundary condition at $r_{N+1}$ is applied.
\section{Numerical results}\label{res}
Here, we study the time and space evolution of the fast solar wind quantities
 ($n,U,T_{\bot p}, T_{\| p}, T_{\bot e}, T_{\| e},\\ q_{\bot p}, q_{\| p},q_{\bot e},
 q_{\| e}, E_{W}$) by applying the two-fluid model in the kinetic theory framework.
 
Using the Bi-Maxwellian distribution function for protons and Kappa-Maxwellian distribution for electrons, the 11 coupled equations are derived. The numerical solution of the 11 coupled equations for different $\kappa$ index ($\kappa$=2, 5, 7, 30)
 is studied.

Figure \ref{fig2} represents spatial variations of the electron and proton number densities (assumed to be equal, $n_{e}=n_{p}=n$) from the Sun to the near Earth. The number density decreases approximately from 10$^{8} $ cm$^{-3}$ close the Sun to 1.67, 1.86, 2.16, and 2.61 cm$^{-3}$ in near the Earth environment for $\kappa$=2, 5, 7, and 30, respectively. Near the Earth, the density increases with increasing $\kappa$ index.
For high $\kappa$ index, the number density is in good agreement with observations recorded by $Ulysses$ and previous studies \citep[e.g.,][]{ref:2011ApJ...743..197C}.\\
Close to the Sun, the density is comparable to the observed data near the solar minimum (\cite{allen}, Table 14.19 therein).
Power-law functions, $n(r)\sim r^{-\alpha}$, with the exponents $\alpha \approx$ (2.224$\% $, 2.194$\% $, 2.154$\% $, and 2.04$\% $) $\pm$ 0.04$\% $ are fitted to the number density at (0.3 - 1) AU for $\kappa = 2, 5, 7,$ and $30$ at (0.3 - 1) AU from the Sun.

Figure \ref{fig3} shows the profiles of the fast solar wind speed $U$ and the Alfv\'en speed $V_{A}$ for different $\kappa$ indices.
Close to the Sun, outflow and Alfv\'en velocities decrease with decreasing $\kappa$. Expectedly, for large $\kappa$ the results are in agreement with the Maxwellian model for electrons \citep{ref:2011ApJ...743..197C}.
The position of the Alfv\'en critical point (at this point the outflow velocity reaches the Alfv\'en velocity) is obtained as $r_{A}/R_{\odot} = 8.5, 8.4, 8.33,$ and $8.31$  for $\kappa=2, 5, 7,$ and $30$, respectively. We obtain the 
 $U(r_{A}) = 601.45, 615.55, 619.18$, and $617.47$ km/s, respectively. The outflow velocities are obtained as  $U(r_{AU}) \approx 822.85, 816.38, 806.36,$ and $804.63$~km/s near the Earth environment. The simulated Alfv\'en velocity near the Earth is in agreement with the observational values ranging from 4.2 to 160.5 km/s at (0.3 - 0.7) AU \citep{ref:1982JGR....87...35M}.

The parallel and perpendicular components of both proton and electron temperatures are demonstrated in Figures \ref{fig4}, and \ref{fig5}, respectively. It is clearly shown that the proton temperature in each direction increases with increasing $\kappa$ index. From the Sun to about $r\approx 26 R{\odot}$ the $T_{\perp p}$ is significantly more than $T_{\parallel p}$ for all $\kappa$. In region $26<r/R_{\odot}<36$ the parallel temperature rises above the perpendicular component, which, is in agreement with \cite{ref:2011ApJ...743..197C}.

Close to the Sun, the two components of electron temperatures are approximately the same. For small $\kappa$ the high million kelvin temperatures for electrons are in good agreement with both observations and a previous study \citep[e.g.,][]{zouganelis2004transonic} at the solar atmosphere. Expectedly, close the Sun the Maxwellian behavior for electrons is obtained for large $\kappa$ \citep[e.g.,][]{ref:2011ApJ...743..197C}.
The difference between the two components of the temperatures increases after about $r/R_{\odot}\approx3.92, 4.06, 4.25,$ and $4.7$  for $\kappa=2, 5, 7,$ and $30$, respectively.

Observations show that $T_{\| e}/T_{\perp e}$ tends to $1.2$ in near the Earth environment, \citep[e.g.,][]{ 1975JGR....80.4181F,1987JGR....92.1103P,vstverak2008electron}. We find this ratio to be about
$T_{\| e}/T_{\perp e}$= 1.1, 1.06, 1.05, and 1.02 for $\kappa=2, 5, 7,$ and $30$, respectively, at 1 AU from the Sun. 
For both components, the exponents of the fitted power-law functions ($T\approx r^{-\alpha}$) at a distance (0.3 - 1) AU are shown in Figure \ref{fig5}.

Figure \ref{fig6} shows the heating rate ratio (the ratio of the turbulent heating rate to the total heating rate) for both components of the electrons ($Q_{\parallel e}/Q, Q_{\perp e}/Q$) and protons ($Q_{\parallel p}/Q,Q_{\perp p}/Q$), and the total heating rate for electrons ($Q_{e}/Q$) for different $\kappa$ indices.  The cures present the behavior of the energy exchanges between the shear Alfv\'en wave (and/or KAW) and the particles.
 As shown in the figure, close the Sun most of the wave-dissipated energy is absorbed by electrons in the parallel direction. Also, the absorbed energy increases with decreasing $\kappa$ index. The absorbed energy by protons ($Q_{\parallel p}/Q, Q_{\perp p}/Q$) and electrons ($Q_{\perp e}/Q$) increases with increasing distance from the Sun $(r)$.  Expectedly, for large $\kappa$ index, both components of proton heating rates and the total heating rate for electrons are in good agreement with the previous study \cite[e.g.][]{ref:2011ApJ...743..197C}. It is shown that the total turbulent heating rates ($Q_{\parallel e}/Q + Q_{\perp e}/Q +Q_{\parallel p}/Q+Q_{\perp p}/Q$) approaches the unity. 

The behaviors of the heat flux components (parallel and perpendicular) for both protons and electrons are represented in Figures \ref{fig7}, and \ref{fig8},  respectively.
Both components of the proton heat flux decrease with decreasing $\kappa$. Near the Earth, the parallel component of proton heat flux is larger than the perpendicular component for all $\kappa$. The free-streaming heat flux for proton is given by (Equation \ref{fsp}),
\begin{eqnarray}\label{fsp}
q_{\rm fs,p}=1.5nk_{B}T_{p}v_{\rm tp},
\end{eqnarray}
where $v_{\rm tp}=\sqrt{k_{B}T_{p}/m_{p}}$.
As we see in the figure, $q_{\rm fs,p}$ decreases with decreasing $\kappa$. It is clearly shown that $q_{\perp p}$ and $q_{\parallel p}$ are smaller than free-streaming heat flux from the Sun to near the Earth.

The perpendicular and parallel component of the electron heat flux increase with decreasing $\kappa$. Generally, the electron heat flux in the perpendicular direction is significantly more than the parallel one for all $\kappa$. 
Also, close to the Sun, the total electron heat flux $(q_{e}=\frac{2q_{\perp e}+q_{\|e}}{3})$ is approximately equal to the Spitzer approximation \citep{spitzer1953transport},
\begin{eqnarray}
q_{\rm sh}= -\kappa_{e0}T_{e}^{5/2}\frac{\partial T_{e}}{\partial r},
\end{eqnarray}
where $\kappa_{0e}=\frac{1.84\times 10^{-5}}{\ln \Lambda}~ {\rm erg~ s^{-1} K^{-7/2} cm^{-1}}$, and $\ln \Lambda$ is the Coulomb logarithm.
Near the Earth, the electron heat flux tends to the electron free-streaming heat flux as
\begin{eqnarray}
q_{\rm fs,e}=1.5nk_{B}T_{e}v_{\rm te},
\end{eqnarray} 
where $v_{\rm te}=\sqrt{k_{B}T_{e}/m_{e}}$, and is comparable with $Helios$ data for electron heat flux at 1 AU \citep{LeChat2012}.
A considerable difference between the parallel and perpendicular components of electron heat flux may  be related to the wave-particle interactions in plasma (Figure \ref{fig6}) and also the non-Maxwellian distribution for electrons. According to Figure \ref{fig6},  the absorbed energy of electrons (in wave-particle interactions) decreases in the parallel direction but increases in the perpendicular direction from the Sun to Earth. Another factor affecting this difference may be related to transporting the electron energy from the parallel to the perpendicular direction \citep{vstverak2015electron}. 
The power exponents $\alpha_{\perp}$ , $\alpha_{\|}$ for the power-law function fitted to the fluxes at (0.3 - 1) AU are presented in Figure \ref{fig8}. 
The value of the electron and proton heat fluxes, temperatures, and solar wind energy fluxes for various $\kappa$ indices at 1 AU are listed in Table \ref{table:nonlin}.\\
\begin{table}[ht!]
	\caption{The Components of Heat Flux for Electrons and Protons, the Components of Temperature for Electrons and Protons, and the Total Energy Flux for Different $\kappa$ (at 1 AU) Are Tabulated.} 
	\centering
	\begin{tabular}{|c | c | c | c | c |} 
		\hline\hline
		$\kappa$ Index & 2 & 5 & 7 & 30 \\ [0.5ex] 
		\hline 
		$q_{\|e}(\rm W/m^{2})$ & $8.3\times10^{-13}$ & $5.86\times10^{-13}$ & $4.45\times10^{-13}$ & $3.3\times10^{-13}$ \\ 
		$q_{\perp e} (\rm W/m^2)$ & $1.12\times10^{-5}$ & $8.3\times10^{-6}$ & $7.5\times10^{-6}$ & $6.99\times10^{-6}$ 
		\\
		$q_{\|p}(\rm W/m^{2})$ & $7.83\times10^{-11}$ & $8.01\times10^{-11}$ & $8.3\times10^{-11}$ & $8.7\times10^{-11}$
		\\
		$q_{\perp p}(\rm W/m^{2})$ & $2.19\times10^{-11}$ & $3.14\times10^{-11}$ & $3.5\times10^{-11}$ & $4.1\times10^{-11}$
		\\
		$T_{\|e}(\rm K)$ & $1.64\times10^{5}$ & $1.32\times10^{5}$ & $1.09\times10^{5}$ & $9.32\times10^{4}$
		\\
		$T_{\perp e}(\rm K)$ & $1.54\times10^{5}$ & $1.25\times10^{5}$ & $1.01\times10^{5}$ & $9.14\times10^{4}$
		\\
		$T_{\|p}(\rm K)$ & $1.6\times10^{5}$ & $1.7\times10^{5}$ & $1.8\times10^{5}$ & $1.9\times10^{5}$
		\\
		$T_{\perp p}(\rm K)$ & $1.16\times10^{5}$ & $1.37\times10^{5}$ & $1.52\times10^{5}$ & $1.6\times10^{5}$
		\\
		$F_{\rm tot}(\rm W/m^{2})$ & $7.8\times10^{-4}$ & $8.5\times10^{-4}$ & $9.5\times10^{-4}$ & $1.1\times10^{-3}$
		\\
		
		\hline 
	\end{tabular}
	\label{table:nonlin} 
\end{table} 
\clearpage
\section{Conclusion}\label{conclu} 
Electrons and protons are the main components of the solar wind, so the two-fluid model in the presence of some kinetic effects is useful to study the characteristics of the system. Observational proofs, such as the anisotropic behavior of the temperature of the solar wind electrons, show that the electrons distribution deviates from the well-known Maxwellian distribution.\\
In this paper, we provided a two-fluid model for the solar wind consisting of the Bi-Maxwellian distribution for protons and the Kappa-Maxwellian distribution for electrons. As the Kappa distribution function might tend to the Maxwellian in the limit of the large $\kappa$ index, the small $\kappa$ (less than $ 5$) showed more deviation from the Maxwellian.\\
We derived 11 coupled equations for fast solar wind model quantities, namely 
$n,U,T_{\bot p}, T_{\| p}, T_{\bot e}$, $T_{\| e}, q_{\bot p}, q_{\| p},q_{\bot e}, q_{\| e}, E_{W}$. To this end, we calculated the velocity space moments up to the fourth order. The  functional forms of the six equations (Equations \ref{eq23}, \ref{eq27}, \ref{eq28}, \ref{Tparp}-\ref{eq33}) have the same form seen in \cite{ref:2011ApJ...743..197C}, which was derived for the Bi-Maxwellian protons.
We also presented five new equations (Equations \ref{eq24}-\ref{qpare}, \ref{Tper}, and \ref{Tpar}) in the presence of the Kappa-Maxwellian distribution for electrons. The $\kappa$ changing from 1.5 to infinity is a characteristic of the Kappa distribution and is considered as a free parameter for the present model.

\begin{figure}[ht!]
	\centering
	\includegraphics[width=18cm]{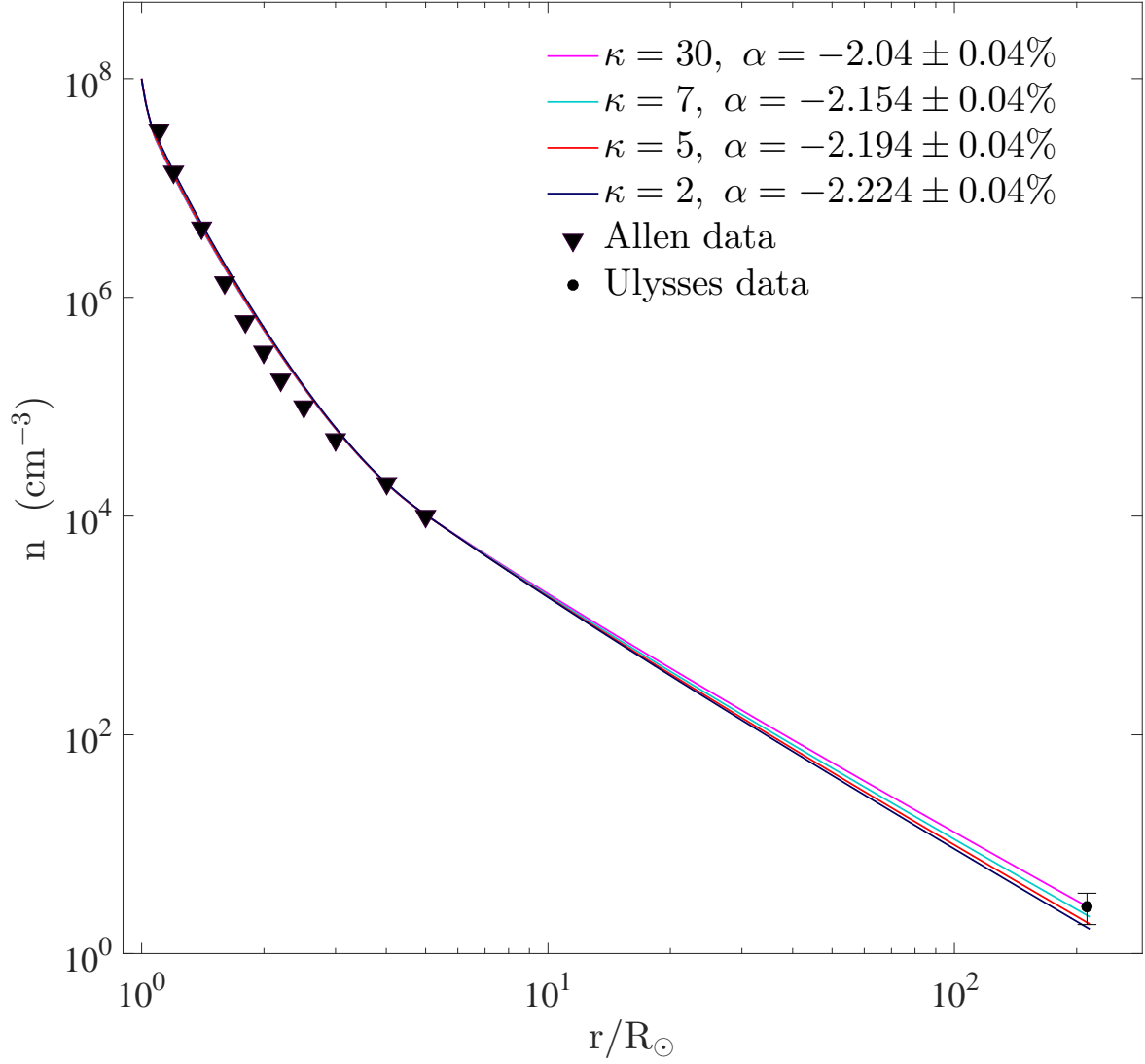}
	\caption{Number density for $\kappa= 2, 5, 7,$ and $30$ vs the distances ($r/R_{\odot}$). The filled circle ($\bullet$) is mean proton density measured by $Ulysses$ at its first orbit \citep{ref:2000JGR...10510419M}. The $(\blacktriangledown s)$ show the polar coronal hole observed data near the solar minimum \citep{allen}. The exponent of the fitted power-law function to the density at (0.3 - 1) AU is presented for all $\kappa$.}\label{fig2}
\end{figure}

\begin{figure}[ht!]
	\centering
	\includegraphics[width=18cm]{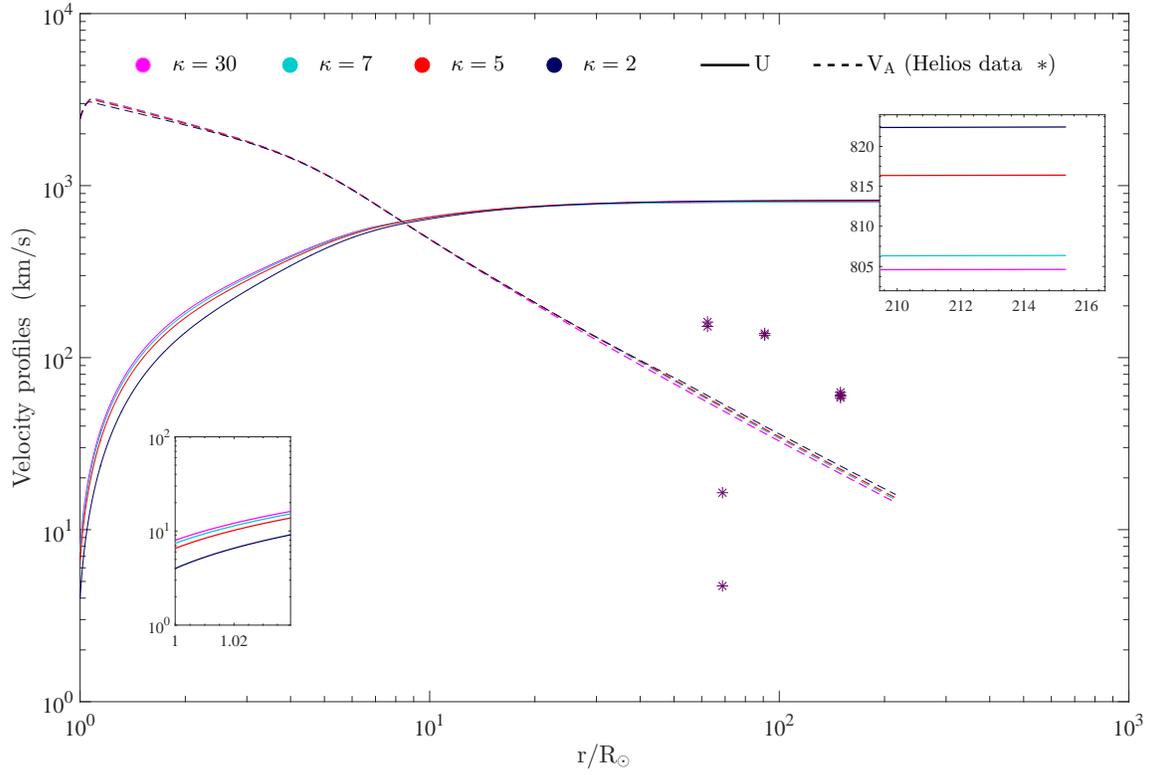}
	\caption{The solid lines are solar wind outflow velocities and the dashed lines are the Alfv\'en velocities for $\kappa= 2, 5, 7,$ and $30$ from the Sun to near the Earth. The $(*s)$ show the $Helios$ data reported for the fast solar wind \citep{ref:1982JGR....87...35M}. 
		The inset box shows the dependency of the velocities on the $\kappa$ index close to the Sun and near the Earth.}\label{fig3}
\end{figure}

\begin{figure}[ht!]
	\centering
	\includegraphics[width=18cm]{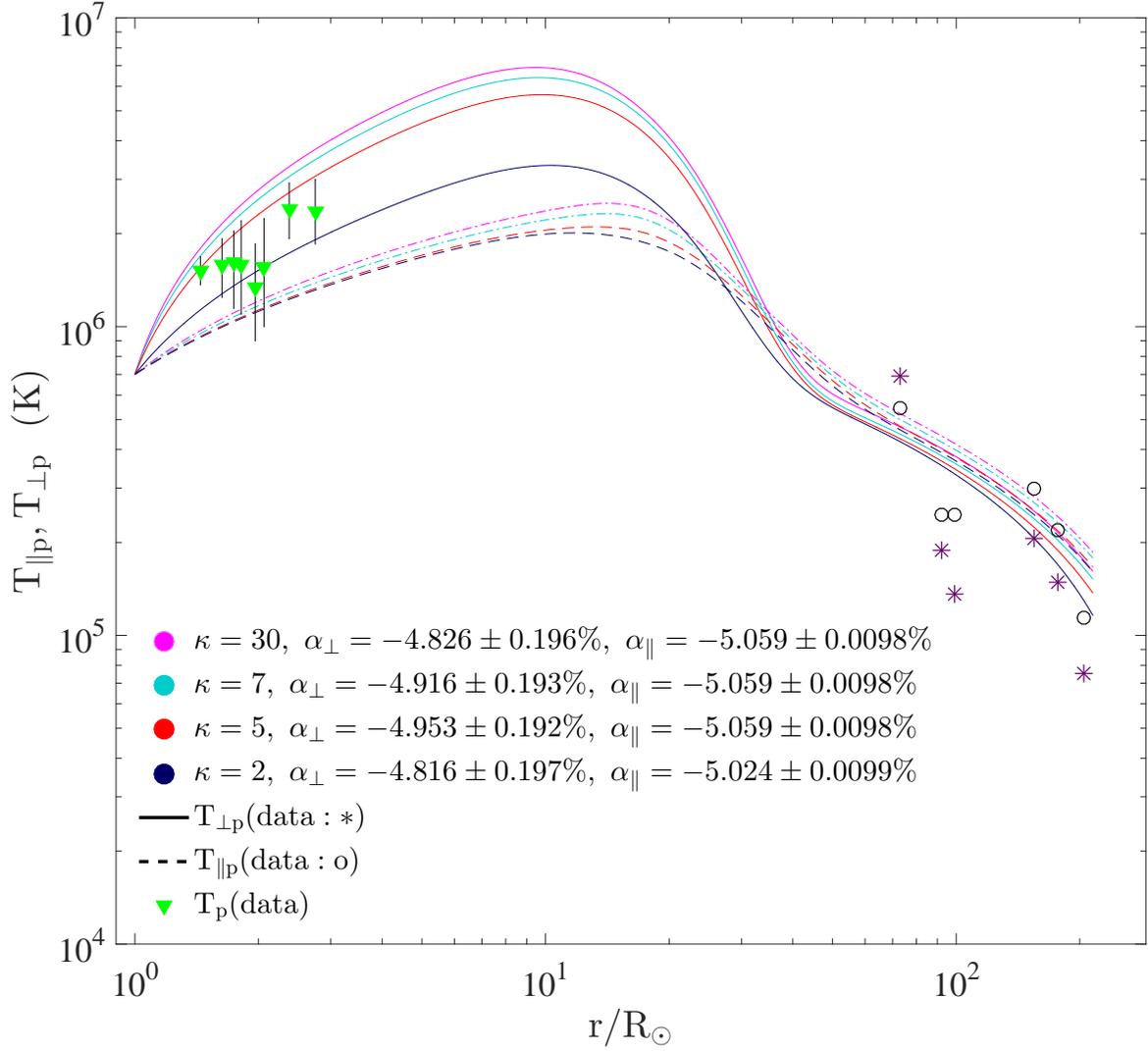}
	\caption{The solid lines represent the perpendicular temperature and the dashed lines represent the parallel temperature for protons for $\kappa= 2, 5, 7,$ and $30$ vs the distance from the Sun. The $(*s)$ and $(\circ s)$ show the $Helios$ data reported for parallel and perpendicular temperatures for the fast solar wind, respectively \citep{ref:1982JGR....87...35M}, the $(\blacktriangledown s)$ show the UVCS/$SoHO$ data for the proton temperature, the power-law indices ($\alpha_{\perp}$ ,$\alpha_{\|}$) for all $\kappa$ are presented.   
	}\label{fig4}
\end{figure}

\begin{figure}
	\centering
	\includegraphics[width=18cm]{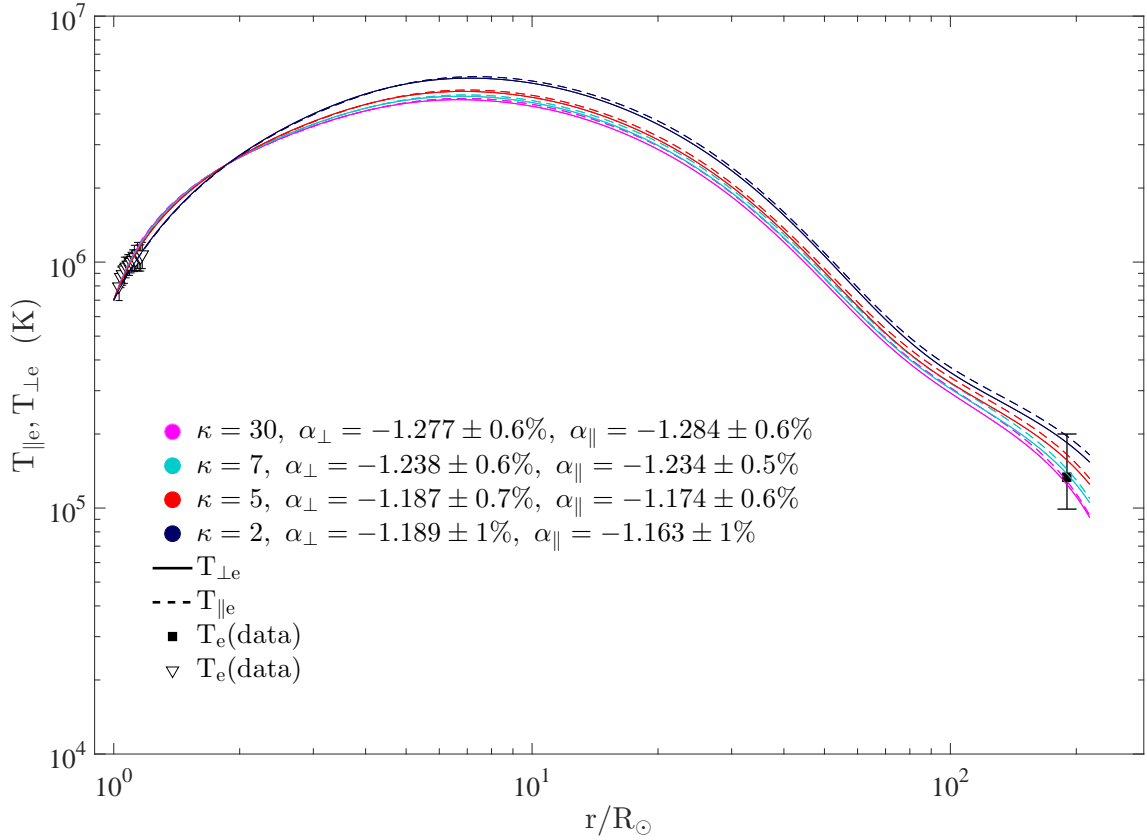}
	\caption{The solid lines are perpendicular temperature and the dashed lines are the parallel temperature for electrons for $\kappa= 2, 5, 7,$ and $30$ from the Sun to the Earth. The $(\triangledown s)$ show the $SoHO$/SUMER data of electron temperature in a polar coronal hole \citep{Landi2008} and the square ($\blacksquare$) is the mean electron temperature for the fast solar wind that measured by ISEE 3 and $Ulysses$ \cite{newbury1998electron}. $\alpha_{\perp}$ and $\alpha_{\|}$ are the power-law indices for $\kappa= 2, 5, 7,$ and $30$ at (0.3 - 1)AU.}\label{fig5}
\end{figure}

\begin{figure}[ht!]
	\centering
	\includegraphics*[width=20cm]{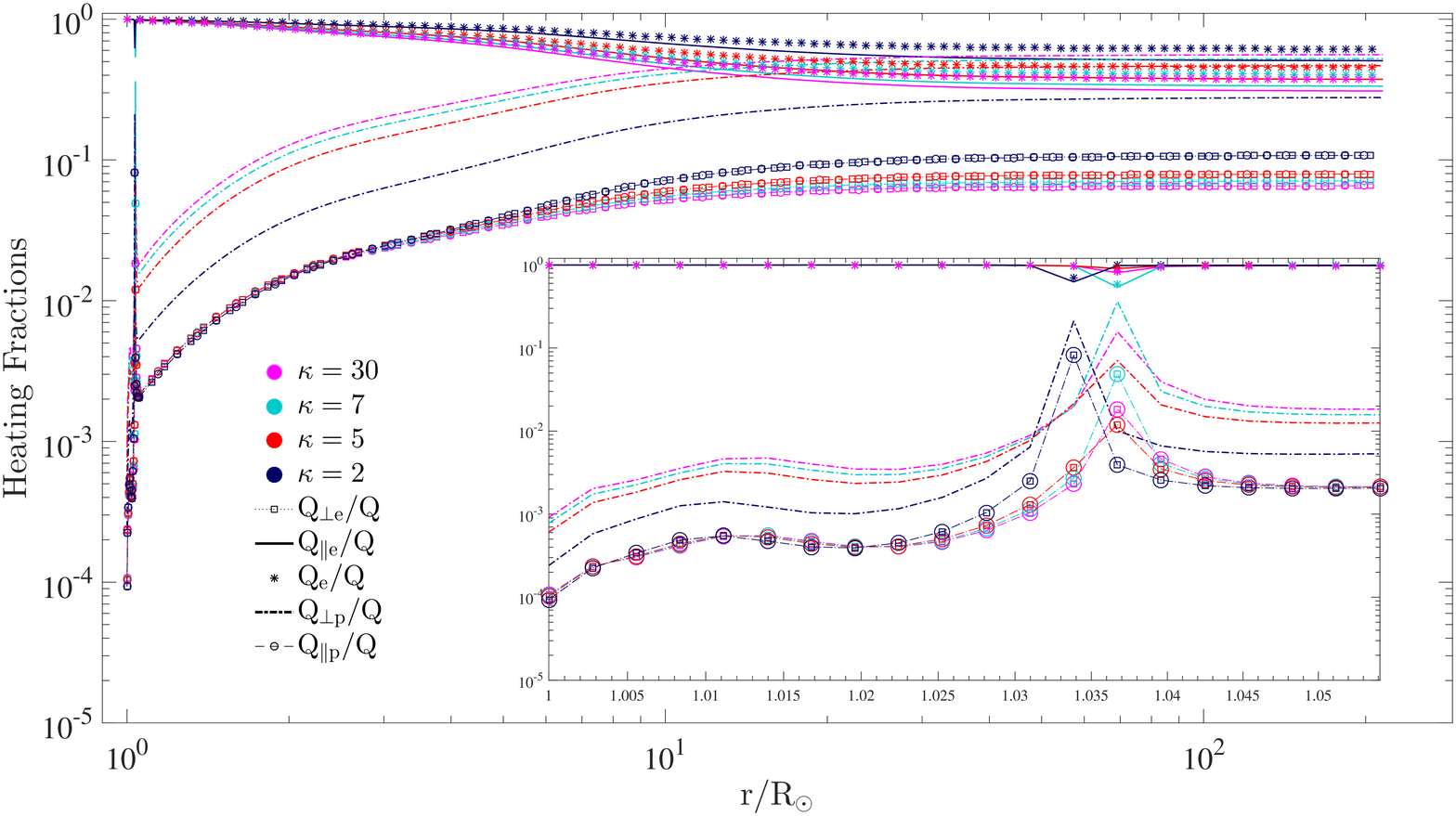}
	\caption{Ratio of the turbulence heating rates of electrons and protons ($\|,\perp$) and total heating rate for electrons for $\kappa= 2, 5, 7,$ and $30$. The inset box shows the variability of the mentioned parameters close to the Sun.}\label{fig6}
\end{figure}

\begin{figure}
	\centering
	\includegraphics[width=18cm]{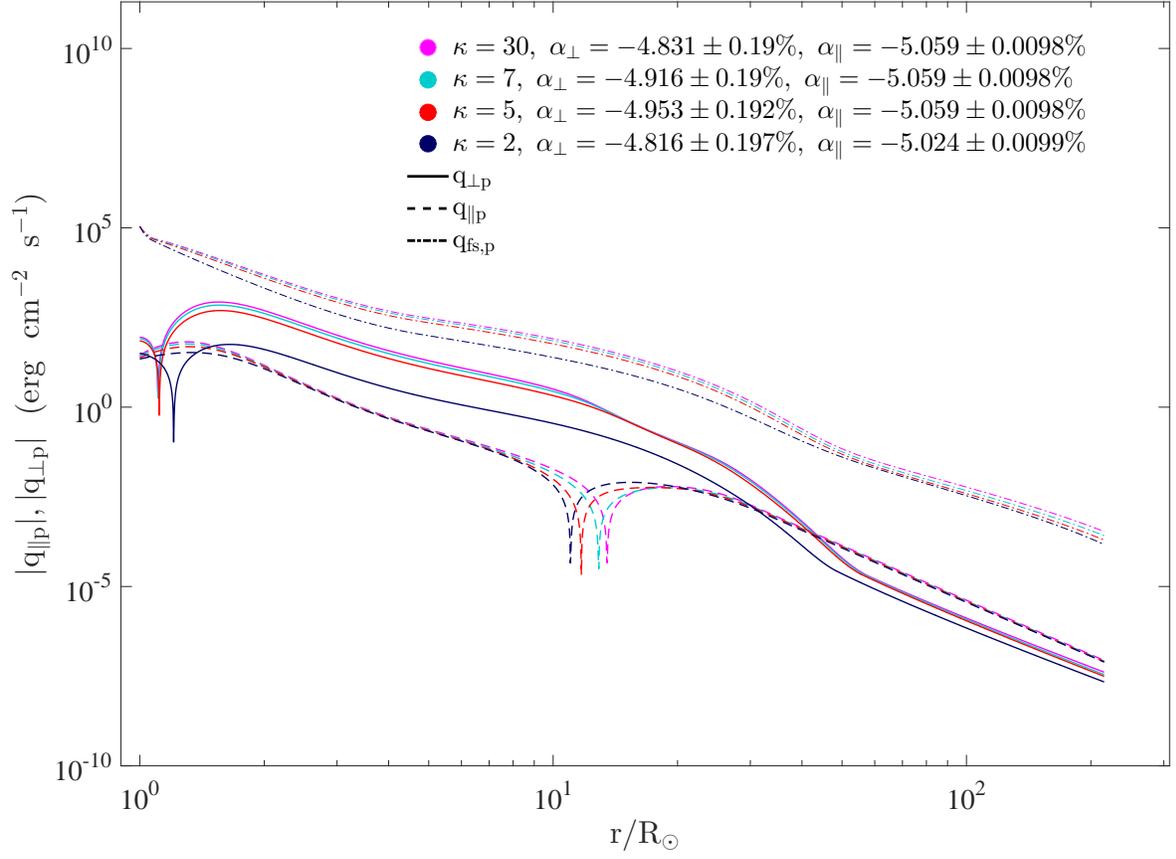}
	\caption{The perpendicular heat flux $(q_{\bot p})$, the parallel heat flux $(q_{\parallel p})$, and the free-streaming analytical calculation $(q_{\rm fs,p})$ of the protons are presented. The power-law indices ($\alpha_{\perp}$ ,$\alpha_{\|}$) for different $\kappa$ are presented.
			}\label{fig7}
\end{figure}

\begin{figure}
	\centering
	\includegraphics[width=18cm]{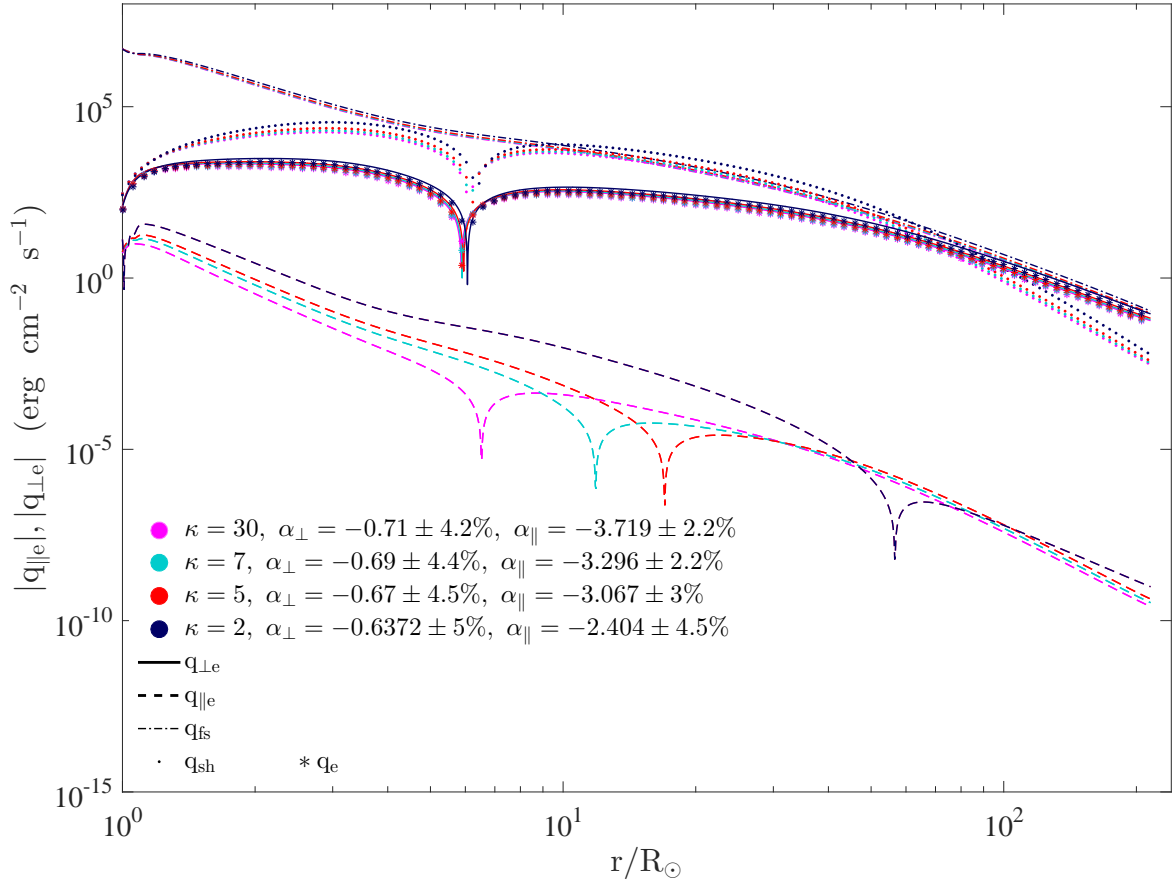}
	\caption{The perpendicular heat flux $q_{\bot e}$, the parallel heat flux $q_{\parallel e}$, the free-streaming analytical calculation $q_{\rm fs,e}$, the Spitzer-H$\ddot{a}$rm $q_{\rm sh}$ heat flux, and the total heat flux $q_e = \frac{2q_{\perp e}+q_{\| e}}{3}$ for the electrons are presented. The power-law exponents for the heat fluxes at (0.3 - 1) AU are obtained.
	}\label{fig8}
\end{figure}

We showed that in the limit of the large $\kappa$, the equation for electron parallel heat flux (Equation \ref{eq28}) behaves like the equation for the proton parallel heat flux (Equation \ref{qpare}), which derived in the presence of the Maxwellian distribution for electrons.
We also used the Landau damping model for the exchange of energies between the particles and waves (shear Alfv\'en wave).\\
Applying the initial and boundary conditions, and the ICN numerical method, the set of equations were solved. The main results are as follows:
\begin{itemize}
	\item[1.]
	 Expectedly, the number density $n(r)$ (assumed to be equal for electrons and protons) shows the scale-free behavior and decreases with increasing the distance from the Sun. The power-law exponent ($\alpha$) for the density at (0.3 - 1) AU was obtained as $\sim$ 2.0 - 2.3 for different $\kappa$, which is in agreement with observations recorded by $Helios$ \citep{vstverak2015electron}. This power-law behavior may be related to the nature of the Kappa distribution. The power-law behavior for the number density was reported in the literature \citep[e.g.,][]{Erickson1964, allen, vstverak2015electron}. Also, near the Earth, the number density decreases with decreasing $\kappa$ index. 
	\item[2.]
	The outflow speed increases with increasing distance from the Sun ($r$), while the Alfv\'en velocity decreases with $r$.
	For small $\kappa$ index, the Alfv\'enic critical point occurs at a distance close to the Sun.
	It seems the solution for solar wind flow is analogous to the properties of plasma flow in the  Alfv\'enic black hole. An Alfv\'enic black hole may be created using a magnetic flux tube with a variable cross section and super-Alfv\'enic flow. Using the linearized MHD equations in the presence of super-Alfv\'enic plasma flow, a tensorial form of the Alfv\'en waves (with an accompanying metric) was obtained \citep{2018EPJC...78..662G}. The resultant metric is singular at a point (horizon of black hole) where the local Alfv\'en speed is equal to flow speed. The horizon of the Alfv\'enic black hole is likely similar to the critical Alfv\'enic point in the solar wind solution. In the solar wind solution, from the lower solar atmosphere (lower corona) the flow starts to accelerate (with considerable acceleration $dU/dr>0$) and at the critical Alfv\'enic point (like to the horizon of the Alfv\'enic black hole), the flow speed equals to the Alfv\'enic speed and continuous to very slightly accelerate and approach approximately constant speed far from the Sun as the flux tube diverges.
	Close to the Sun, the fast solar wind propagates with low speeds for small $\kappa$ (Figure \ref{fig2}). But near the Earth, high speed is related to the small $\kappa$.
	We found the outflow speed is in the range of 804-822 km/s (near the Earth) and satisfies with the observational data \citep{bame1993ulysses,feldman2005sources}. Expectedly, for large $\kappa$, the value of the outflow speed is in good agreement with \cite{ref:2011ApJ...743..197C}, who they used a Maxwellian distribution for electrons.
	\item[3.]
	Close to the Sun and for $\kappa$ around 7, the proton temperature is in consistent with the observational value (Figure \ref{fig4}). This is also in agreement with \cite{2016SoPh..291.2165P}, who modeled the electrons close to the Sun.
	The parallel and perpendicular components of the electron temperature for $\kappa$=7 are also comparable with observations. Getting away from the Sun, small $\kappa$ shows temperatures of several million kelvin for electrons \citep{zouganelis2004transonic}. This high temperature is related to the extraordinary nature of the solar atmosphere (corona). Near the Earth and for small $\kappa$, the value of the electron temperature ratio ($T_{\|e}/T_{\perp e}$) was obtained as about 1.1, which is in agreement with observationas \citep{vstverak2008electron}.
	 \item[4.]
	The perpendicular and parallel heat flux components of electrons increase with decreasing $\kappa$ index. The electron heat flux approximately is comparable with the free-streaming analytical curve (collisionless regime) near the Earth and the Spitzer-H$\ddot{a}$rm solution (Collisional regime) near the Sun.
		 
\end{itemize}

Finally, this study shows that while some of the observational quantities (e.g., electron and proton temperature) are well modeled with $\kappa$=7 close to the Sun and far away from the Sun (near the Earth), other quantities ({\rm e.g.,} the temperature ratio for electrons) are satisfied with a small $\kappa$ index (less than 5). Thus, this study encourages us to develop the multi-index models including three or more $\kappa$ indices ( large $\kappa$ close the Sun and small near the Earth) for the fast solar wind.



\clearpage	
\section*{Acknowledgement}
The authors would like to thank  the $Helios$ team for the solar wind data. We also thank Dr. Danial Verscharen
University College London-Mullard Space Science Laboratory for his very helpful discussion. We would like to give a special thanks to Professor Stefaan Poedts for his suggestion regarding the idea of the effects of Kappa distribution for electrons in a two-fluid model for solar winds. 
 We also gratefully thank the anonymous referee for very helpful and constructive comments and suggestions that improved the manuscript. The authors would like to express their gratitude to the Iran National Science Foundation (INSF) for supporting this research under grant No. 93043701. We acknowledge the support from the High Performance Computing Center of Department of physics, University of Zanjan.

\bibliography{refs}
\end{document}